\documentclass[final,5p,times,twocolumn]{elsarticle}

\usepackage{amsmath}    
\usepackage{amssymb}    
\usepackage{booktabs}   
\usepackage{microtype}  
\usepackage{hyperref}   
\usepackage{cleveref}   
\usepackage[table]{xcolor}
\usepackage[most]{tcolorbox}
\usepackage[scaled=0.9]{inconsolata}
\usepackage{dashrule}

\hypersetup{
    colorlinks=true,
    linkcolor=blue,
    filecolor=magenta,      
    urlcolor=cyan,
    citecolor=blue,
}

\definecolor{lightpurple}{RGB}{250,248,255}

\newtcolorbox{promptbox}{
    colback=lightpurple,
    colframe=black,
    boxrule=0.6pt,
    arc=2pt,
    left=6pt,
    right=6pt,
    top=6pt,
    bottom=6pt,
    width=\columnwidth,
    fontupper=\ttfamily\small
}


\begin{document}

\begin{frontmatter}

\title{Wave2Word: A Multimodal Transformer Framework for Joint EEG-Text Alignment and Multi-Task Representation Learning in Neurocritical Care}






\author[inst1]{Argha Kamal Samanta\fnref{fn1}\corref{cor1}}
\ead{arghakamal25@gmail.com}

\author[inst2]{Deepak Mewada\fnref{fn1}}
\ead{deepakmewada96@gmail.com}

\author[inst3]{Monalisa Sarma}
\ead{monalisa@iitkgp.ac.in}

\author[inst2]{Debasis Samanta}
\ead{dsamanta@iitkgp.ac.in}

\fntext[fn1]{These authors contributed equally to this work.}

\affiliation[inst1]{organization={Department of Electronics and Electrical Communication Engineering},
            addressline={Indian Institute of Technology, Kharagpur}, 
            city={Kharagpur},
            postcode={721302},
            state={West Bengal},
            country={India}}
\affiliation[inst2]{organization={Department of Computer Science and Engineering},
            addressline={Indian Institute of Technology, Kharagpur}, 
            city={Kharagpur},
            postcode={721302},
            state={West Bengal},
            country={India}}

\affiliation[inst3]{organization={Subir Chowdhury School of Quality and Reliability},
            addressline={Indian Institute of Technology, Kharagpur}, 
            city={Kharagpur},
            postcode={721302},
            state={West Bengal},
            country={India}}

\begin{abstract}
Continuous electroencephalography (EEG) is routinely used in neurocritical care to monitor seizures and other harmful brain activity, including rhythmic and periodic patterns that are prevalent and clinically significant. Although deep learning methods have achieved high accuracy for seizure detection, most existing approaches remain seizure-centric, rely on discrete-label supervision, and are evaluated primarily using accuracy-based metrics.
A central limitation of current EEG modeling practice is the weak correspondence between learned representations and how EEG findings are interpreted and summarized in clinical workflows. Harmful EEG activity exhibits overlapping patterns, graded expert agreement, and temporal persistence, which are not well captured by classification objectives alone
This work proposes a multimodal EEG representation learning framework that integrates signal-domain modeling with structured clinical language supervision. First, raw EEG is transformed into a longitudinal bipolar montage and time--frequency representations. Second, dual transformer-based encoders model complementary temporal-centric and frequency-centric dependencies and are fused using an adaptive gating mechanism. Third, EEG embeddings are aligned with structured expert consensus descriptions through a contrastive objective. Finally, an EEG-conditioned text reconstruction loss is introduced as a representation-level constraint alongside standard classification loss.
Experimental evaluation using a controlled train--validation--test split achieves a six-class test accuracy of 0.9797. Ablation analyses show that removing contrastive alignment reduces cross-modal retrieval performance from Recall@10 of 0.3390 to 0.0045, despite minimal change in classification accuracy.
These findings demonstrate that discriminative accuracy does not reliably reflect representation quality for clinically meaningful EEG modeling. Explicit alignment with structured clinical language exposes differences in semantic organization and optimization behavior that are invisible to accuracy-based evaluation, providing a principled basis for assessing EEG representations beyond classification performance.
\end{abstract}



\begin{keyword}
Electroencephalography \sep Neurocritical care\sep Harmful brain activity \sep Multimodal learning \sep EEG--language alignment 
\end{keyword}

\end{frontmatter}

\section{Introduction}
\label{sec:introduction}

In modern intensive care units, continuous electroencephalography (EEG) is routinely used to monitor patients with acute brain injury, coma, or unexplained altered mental status, where clinical examination alone is insufficient. Large multicenter and prospective studies report electrographic seizures in approximately 10--40\% of critically ill adults undergoing continuous EEG monitoring, with more than 80\% of these seizures being non-convulsive and therefore clinically silent \cite{Claassen2004,Claassen2013,Westover2015,RodriguezRuiz2017}. Beyond seizures, rhythmic and periodic EEG patterns, including lateralized periodic discharges (LPDs), generalized periodic discharges (GPDs), lateralized rhythmic delta activity (LRDA), and generalized rhythmic delta activity (GRDA), are even more prevalent. Multiple ICU cohorts report these patterns in more than half of monitored patients \cite{Foreman2012,Gaspard2014,Saab2020,Abend2013}.

These EEG patterns are clinically significant. Prior studies demonstrate independent associations between rhythmic or periodic EEG activity and increased mortality, prolonged intensive care unit length of stay, delayed neurological recovery, and poor functional outcomes \cite{Gaspard2014,Westover2015,Struck2017,Young2018,Payne2019}. Adverse outcomes correlate more strongly with pattern burden, persistence, and temporal evolution than with the presence of a single categorical abnormality \cite{RodriguezRuiz2017,Payne2014,Newey2020}. At the same time, continuous EEG monitoring commonly extends for 24--72 hours or longer per patient, creating a substantial interpretive burden that requires sustained expert review \cite{Pandian2004,Abend2013}. These factors motivate automation that supports clinical interpretation rather than replacing it.

A key challenge in this setting is that harmful EEG activity does not form a set of discrete and mutually exclusive classes. Clinical neurophysiology literature describes seizures, rhythmic activity, and periodic discharges as occupying an ictal--interictal continuum, characterized by gradual transitions, overlapping morphologies, and temporal evolution \cite{Foreman2012,Hirsch2013,Struck2017}. EEG patterns may coexist within the same recording or transition between rhythmic, periodic, and ictal states over time \cite{RodriguezRuiz2017,Payne2019}. As a result, expert interpretation often depends on relative dominance, persistence, and temporal context rather than binary presence or absence.

This complexity is reflected in expert agreement. Inter-rater agreement for rhythmic and periodic EEG patterns is only moderate even among experienced neurophysiologists, with reported Cohen’s kappa values typically ranging from 0.4 to 0.6 \cite{Gerber2008,Gaspard2014,Halford2015}. To address this variability, the American Clinical Neurophysiology Society (ACNS) introduced standardized terminology that defines harmful EEG patterns while explicitly encoding dominance, uncertainty, and expert disagreement \cite{Hirsch2013}. In routine clinical practice, EEG findings are summarized using percentage agreement across experts rather than definitive labels, reflecting the probabilistic nature of interpretation \cite{Beniczky2020,Halford2015}. Computational formulations that ignore this structure diverge from clinical reporting conventions.

Machine learning methods for automated EEG analysis have advanced rapidly in recent years. Convolutional neural networks, temporal convolutional models, and transformer-based architectures achieve strong performance in seizure detection, sleep staging, and general EEG classification \cite{Schirrmeister2017,Lawhern2018,Roy2019,Craik2019,Song2022NeuroImage}. However, systematic reviews show that most studies remain seizure-centric or rely on binary abnormal--normal formulations \cite{Roy2019,Shoeibi2021}. Non-seizure harmful patterns such as LPDs, GPDs, LRDA, and GRDA are often merged into coarse abnormal categories or excluded from model outputs \cite{TjepkemaCloostermans2020,Mousavi2021TBME}.

This modeling choice introduces a mismatch with clinical EEG practice. Discrete-label optimization assumes sharp class boundaries, whereas clinicians reason about mixtures of patterns, graded confidence, and temporal evolution \cite{Foreman2012,Hirsch2013}. As a result, models optimized solely for classification accuracy may learn representations that are discriminative but insufficient for report-level interpretation or uncertainty communication.

More broadly, many existing EEG learning frameworks rely on simplifying assumptions that conflict with clinical workflows. EEG abnormalities are treated as discrete entities despite evidence of continuity. EEG is modeled as a unimodal signal-processing problem even though interpretation is externalized through structured language. Supervision is assumed to be label-complete despite expert disagreement being intrinsic and explicitly encoded in clinical terminology. These assumptions suppress uncertainty and limit insight into representational failure modes, while favoring benchmark performance over clinical alignment.

This disconnect is most apparent in how EEG interpretation is communicated. In clinical practice, EEG findings are conveyed through written reports that describe dominant and secondary patterns, temporal context, and expert agreement using standardized language rather than numeric scores \cite{Hirsch2013,Tatum2016}. Language enables clinicians to express uncertainty, communicate probabilistic judgments, and integrate EEG findings with broader clinical context \cite{Halford2015,Beniczky2020}. Automated systems that do not support or preserve this descriptive structure fail to model the final output of EEG interpretation.

Recent advances in multimodal representation learning suggest a possible direction. Aligning high-dimensional signals with language has been shown to induce semantically structured embeddings in vision and medical imaging domains \cite{Radford2021,Jia2021,Irvin2019,Boecking2022,Huang2023}. Foundation models integrating language with biomedical data further demonstrate that linguistic supervision can regularize representations beyond label-only training \cite{Alsentzer2019,Moor2023,Hong2023}. Despite these advances, EEG has received limited attention in this paradigm, even though EEG interpretation is inherently language-mediated.

Motivated by these observations, this work examines whether aligning EEG representations with structured expert clinical language provides a useful constraint for modeling harmful EEG activity. The central premise is that representations supporting both discrimination and clinically coherent description may better reflect how EEG is interpreted and documented in practice. From this perspective, language alignment is treated as a representational probe rather than an end application.

\paragraph{Overview of the proposed approach.}
The proposed framework mirrors key steps of clinical EEG review. Raw scalp EEG is first transformed into a longitudinal bipolar montage to reduce reference-dependent artifacts and emphasize local voltage gradients. A channel-wise short-time Fourier transform captures rhythmic and spectral structure relevant to ACNS-defined patterns. Dual transformer-based encoders model complementary temporal-centric and frequency-centric dependencies, enabling long-range contextual integration. EEG embeddings are aligned with structured expert clinical descriptions using a symmetric contrastive objective, encouraging organization according to clinical semantics rather than label identity alone. An auxiliary EEG-conditioned text decoder reconstructs expert-style summaries and serves as a representation-level consistency constraint.

\paragraph{Illustrative report-level outputs.} The implications of this formulation are illustrated by the following examples, which compare expert consensus descriptions with model-generated summaries:

\begin{promptbox}
\textbf{SAMPLE01}\\
\textbf{Ground Truth:} Expert opinions show complete agreement, with all identifying lateralized periodic discharges (100\%).

\vspace{4pt}
\textbf{Generated:} expert opinions show complete agreement, with all identifying lateralized periodic discharges ( 100 \% ).
\vspace{4pt}
\hdashrule{\columnwidth}{0.4pt}{2pt}
\vspace{4pt}
\textbf{SAMPLE02}\\
\textbf{Ground Truth:} Expert opinions show mixed agreement, with 8\% identifying lateralized rhythmic delta activity, 54\% identifying generalized rhythmic delta activity, 38\% identifying other patterns.

\vspace{4pt}
\textbf{Generated:} expert opinions show mixed agreement, with 8 \% identifying lateralized rhythmic delta activity, 54 \% identifying generalized rhythmic delta activity, 38 \% identifying other patterns.

\end{promptbox}
Here, text generation is not treated as an end application but as a stringent probe that exposes whether EEG embeddings preserve clinically relevant semantics beyond class identity.

\paragraph{Contributions.}
The contributions of this work are summarized as follows:
\begin{enumerate}
    \item A clinically grounded framework for modeling six ACNS-defined harmful EEG activity classes is presented, addressing the seizure-centric focus of prior work.
    \item A multimodal EEG--text representation learning approach is introduced to align EEG embeddings with structured expert clinical descriptions.
    \item A joint optimization strategy combines classification, contrastive alignment, and EEG-conditioned text reconstruction to constrain representation learning.
    \item Representation quality is evaluated using accuracy, cross-modal alignment, architectural ablation, and optimization stability under subject-exclusive protocols.
\end{enumerate}

\paragraph{Organization of the paper.}
Section~\ref{sec:related_work} reviews related literature. Section~\ref{sec:methodology} describes the proposed framework. Section~\ref{sec:experiments} presents experimental results. Section~\ref{sec:discussion} discusses implications and limitations.

\section{Related Work}
\label{sec:related_work}

Automated electroencephalography (EEG) analysis has been studied extensively, motivated by the need to process long-duration recordings in settings where expert interpretation is time-intensive and resource-limited. This section reviews prior work relevant to neurocritical care EEG, with emphasis on how existing methods represent EEG, define supervision, and evaluate performance. The discussion focuses on concrete modeling practices rather than restating clinical motivation.

\paragraph{Automated EEG analysis in neurocritical care.}
Early automated EEG systems relied on handcrafted features derived from time-domain statistics, spectral power, coherence measures, entropy, and heuristic rules \cite{Gotman1982,Wilson2004,Navakatikyan2006}. These approaches were typically designed for narrow objectives such as seizure detection, burst suppression identification, or background abnormality screening. Performance depended strongly on expert-defined thresholds and was sensitive to noise, electrode configuration, and inter-subject variability, which limited robustness in heterogeneous ICU environments \cite{}.

Large observational and prospective studies later established that electrographic seizures, periodic discharges, and rhythmic EEG patterns are prevalent in critically ill patients and are associated with adverse neurological outcomes when recognition or treatment is delayed \cite{Claassen2004,Foreman2012,Gaspard2014,Westover2015,Saab2020}. These findings motivated the development of automated EEG monitoring systems in the ICU. However, most automated approaches studied or deployed in this setting focus on seizure detection or binary abnormal--normal discrimination. Clinically defined non-seizure patterns such as lateralized or generalized periodic discharges and rhythmic delta activity are often merged into coarse categories or excluded from model outputs \cite{TjepkemaCloostermans2020,Mousavi2021TBME}.

\paragraph{Convolutional neural networks and local-context modeling.}
The introduction of deep learning shifted EEG analysis from handcrafted features to data-driven representation learning. Convolutional neural networks (CNNs) demonstrated that discriminative EEG representations can be learned directly from raw or minimally processed signals. Architectures such as DeepConvNet and ShallowConvNet established strong baselines for general EEG decoding tasks \cite{Schirrmeister2017}, while EEGNet emphasized parameter efficiency and suitability for low-resource settings \cite{Lawhern2018}.

CNN-based models have been widely applied to seizure detection, sleep staging, and brain--computer interface tasks \cite{Antoniades2017,Acharya2018,Waytowich2018,Shoeibi2021}. Their defining characteristic is the use of fixed local receptive fields. While effective for short-duration events, this design limits modeling of EEG phenomena defined by persistence, regularity, and temporal evolution over extended time windows, such as periodic discharges or sustained rhythmic activity \cite{Foreman2012,Struck2017,Payne2019}. Consequently, most CNN-based studies operate on short segments with discrete labels and do not explicitly model long-range temporal structure across clinically relevant durations.

\paragraph{Transformer-based models and long-context EEG encoding.}
Transformer architectures have been introduced to address the limited temporal context of convolutional models. Several studies report improved performance in EEG-based emotion recognition, sleep staging, and general classification using transformer encoders or hybrid CNN--Transformer architectures \cite{Song2022NeuroImage,Yuan2023TBME,Chen2023IEEEJBHI}. Self-attention enables flexible integration of information across time, which is advantageous for non-stationary EEG signals.

Despite this representational capability, transformer-based EEG models remain largely unimodal and are typically trained using discrete classification objectives. Their application to neurocritical care EEG is limited, and clinically defined pattern classes such as periodic discharges or rhythmic delta activity are rarely modeled explicitly. Evaluation is dominated by accuracy or area-under-curve metrics, with limited analysis of representation structure, robustness, or alignment with clinical interpretation \cite{Roy2019,Craik2019,Shoeibi2021}.

\paragraph{Time--frequency representations and multi-view fusion.}
Time--frequency analysis, including short-time Fourier transforms and wavelet decompositions, is widely used to characterize oscillatory structure and transient events in EEG \cite{Oppenheim1999,Brunner2014}. Deep learning models operating on spectrograms or combining time-domain and frequency-domain inputs have demonstrated improved performance across several EEG tasks \cite{Bashivan2016,Acharya2018,Hu2021NeuroImage}.

Multi-view and multi-branch architectures extend this idea by fusing complementary spatial, temporal, and spectral representations \cite{Zhou2021NeuroImage,Wang2022TBME,Ding2023TSception}. These approaches typically rely on static fusion strategies such as concatenation or fixed weighting. Such designs do not adapt representation dominance to pattern-specific signal characteristics, despite clinical evidence that seizures, periodic discharges, and rhythmic activity differ fundamentally in their temporal and spectral organization \cite{Foreman2012,Hirsch2013}.

\paragraph{Multimodal learning and EEG--language integration.}
Multimodal representation learning has become central to medical artificial intelligence, particularly through image--text models in radiology and pathology. Language-aligned representations support report generation, cross-modal retrieval, and structured interpretation \cite{Irvin2019,Boecking2022,Huang2023}. Contrastive learning frameworks such as CLIP demonstrate that alignment with language induces semantically organized embeddings beyond label-only supervision \cite{Radford2021,Jia2021}.

Language-aligned foundation models further show that linguistic supervision can regularize representations across biomedical domains \cite{Alsentzer2019,Moor2023,Hong2023}. In contrast, EEG has received limited attention in this paradigm. Existing EEG--text studies are small in scale, rely on weak or post hoc supervision, and rarely use standardized clinical terminology \cite{Zhang2022MICCAI,Liu2023Sensors}. Explicit alignment between EEG representations and expert clinical language remains uncommon in automated EEG research.

\paragraph{Coverage of ACNS-defined EEG pattern classes.}
Across the EEG machine learning literature, there is a clear imbalance in class coverage. Seizure detection dominates published work, while other clinically important patterns defined by ACNS terminology, including lateralized and generalized periodic discharges and rhythmic delta activity, are underrepresented \cite{Hirsch2013,TjepkemaCloostermans2020}. When included, these patterns are often collapsed into broad abnormal categories or treated as secondary labels \cite{Mousavi2021TBME}. This limited coverage restricts insight into how modern architectures represent different harmful EEG patterns and how representational choices affect separability in clinically relevant scenarios.

\paragraph{Evaluation practices and representation-level analysis.}
Evaluation in EEG deep learning studies is dominated by point-estimate metrics such as accuracy and area under the curve. Multiple reviews document persistent issues including subject leakage, limited reporting of training dynamics, and insufficient robustness analysis \cite{Roy2019,Craik2019,Shoeibi2021}. Similar concerns have been raised more broadly in medical artificial intelligence regarding the limitations of accuracy-centric evaluation for clinical deployment \cite{Kelly2019,Roberts2021}.

Few EEG studies examine whether learned representations align with clinical semantics, support uncertainty communication, or remain stable under architectural perturbations. Representation-level analysis and cross-modal evaluation remain uncommon despite their relevance for translation to clinical workflows.

\paragraph{Summary.}
Table~\ref{tab:related_work_summary} summarizes representative prior work and highlights recurring methodological characteristics, including seizure-centric task definitions, unimodal signal modeling, discrete-label supervision, static feature fusion, and accuracy-focused evaluation. While substantial progress has been made in automated EEG classification, open questions remain regarding semantic alignment, class coverage, and representation quality for clinically defined harmful EEG activity.

\begin{table*}[!t]
\centering
\caption{Representative prior work in automated EEG analysis and related multimodal learning, organized by clinical scope, EEG representation, supervision paradigm, and unresolved methodological gaps relative to neurocritical care EEG interpretation.}
\label{tab:related_work_summary}
\resizebox{\textwidth}{!}{
\begin{tabular}{p{5.2cm} p{2.8cm} p{3.4cm} p{3.6cm} p{6.4cm}}
\toprule
\textbf{Work (Year, Venue)} &
\textbf{Clinical Scope} &
\textbf{EEG Representation} &
\textbf{Learning Paradigm} &
\textbf{Observed Methodological Limitations} \\
\midrule

\cite{Gotman1982} (1982, \textit{Electroencephalogr Clin Neurophysiol}) &
Seizure detection &
Handcrafted temporal features &
Rule-based detection &
Narrow task scope; extensive expert tuning; limited robustness across patients \\

\cite{Wilson2004} (2004, \textit{Clin Neurophysiol}) &
ICU EEG screening &
Spectral features with heuristics &
Rule-based system &
Sensitive to montage and noise; no structured EEG pattern taxonomy \\

\cite{Navakatikyan2006} (2006, \textit{IEEE TBME}) &
EEG abnormality detection &
Time--frequency features &
Statistical learning &
Feature engineering dependent; limited cross-dataset generalization \\

\cite{Claassen2004} (2004, \textit{Neurology}) &
ICU seizure burden &
Clinical EEG review &
Observational study &
Establishes prevalence and outcomes; no automated modeling \\

\cite{Foreman2012} (2012, \textit{J Clin Neurophysiol}) &
Ictal--interictal continuum &
Clinical EEG morphology &
Conceptual / observational &
Demonstrates continuity of EEG patterns; not addressed in ML formulations \\

\cite{Schirrmeister2017} (2017, \textit{Hum Brain Mapp}) &
General EEG decoding &
Raw EEG, CNN (DeepConvNet, ShallowConvNet) &
Supervised classification &
Local receptive fields; short temporal context; non-ICU task focus \\

\cite{Lawhern2018} (2018, \textit{J Neural Eng}) &
BCI, general EEG &
Compact CNN (EEGNet) &
Supervised classification &
Accuracy-centric; unimodal; no uncertainty or semantic modeling \\

\cite{Antoniades2017} (2017, \textit{Clin Neurophysiol}) &
Neonatal seizure detection &
CNN-based classifier &
Supervised classification &
Seizure-only focus; dataset-specific validation \\

\cite{Acharya2018} (2018, \textit{Comput Biol Med}) &
Seizure detection &
STFT + CNN &
Supervised classification &
Narrow task framing; excludes non-seizure EEG patterns \\

\cite{TjepkemaCloostermans2020} (2020, \textit{Clin Neurophysiol}) &
ICU EEG abnormality &
Deep CNN &
Binary abnormal/normal &
Collapses ACNS-defined EEG patterns into coarse labels \\

\cite{Mousavi2021TBME} (2021, \textit{IEEE TBME}) &
ICU EEG monitoring &
CNN-based deep learning &
Binary classification &
No differentiation of rhythmic or periodic EEG patterns \\

\cite{Craik2019} (2019, \textit{J Neural Eng}) &
Seizure detection &
CNN/RNN hybrids &
Supervised classification &
Limited modeling of long-range temporal evolution \\

\cite{Roy2019} (2019, \textit{IEEE TNNLS}) &
EEG deep learning survey &
Review &
Systematic analysis &
Identifies subject leakage, limited robustness analysis, weak clinical grounding \\

\cite{Song2022NeuroImage} (2022, \textit{NeuroImage}) &
Emotion recognition &
Transformer encoder &
Supervised classification &
Non-clinical task; unimodal; accuracy-only evaluation \\

\cite{Yuan2023TBME} (2023, \textit{IEEE TBME}) &
EEG pattern recognition &
Transformer-based architecture &
Supervised classification &
Non-ICU dataset; no ACNS pattern grounding \\

\cite{Zhou2021NeuroImage} (2021, \textit{NeuroImage}) &
EEG classification &
Multi-view CNN &
Static feature fusion &
Fusion not pattern-adaptive \\

\cite{Wang2022TBME} (2022, \textit{IEEE TBME}) &
Brain state classification &
Multi-branch deep learning &
Supervised classification &
Heuristic fusion; no semantic or language supervision \\

\cite{Ding2023TSception} (2023, \textit{IEEE TNNLS}) &
General EEG classification &
Temporal--spectral CNN (TSception) &
Supervised classification &
Improved feature diversity; still unimodal and accuracy-centric \\

\cite{Zhang2022MICCAI} (2022, \textit{MICCAI}) &
EEG--text representation learning &
Multimodal deep learning &
Weak alignment &
Limited scale; no ACNS terminology; qualitative evaluation only \\

\cite{Irvin2019} (2019, \textit{AAAI}) &
Radiology reporting &
Image--text alignment &
Multimodal learning &
Demonstrates report-level modeling value (non-EEG domain) \\

\cite{Radford2021} (2021, \textit{ICML}) &
Vision--language (general) &
Contrastive alignment (CLIP) &
Multimodal contrastive learning &
Not applied to EEG; motivates signal--language alignment paradigm \\

\cite{Alsentzer2019} (2019, \textit{ACL}) &
Clinical NLP &
Bio-ClinicalBERT &
Language pretraining &
Does not integrate physiological signals \\

\cite{Moor2023} (2023, \textit{Nature}) &
Medical foundation models &
Multimodal representations &
Language-aligned learning &
EEG underrepresented; primary focus on imaging and EHR \\

\cite{Hong2023} (2023, \textit{Nature}) &
Multimodal clinical AI &
Foundation model paradigm &
Language-centric supervision &
Limited application to EEG or neurocritical care \\

\bottomrule
\end{tabular}
}
\end{table*}

\section{Proposed Methodology}
\label{sec:methodology}

The proposed methodology is designed to learn representations of harmful brain activity that are discriminative for classification while remaining aligned with how EEG findings are summarized using standardized clinical terminology. In this work, alignment is defined with respect to structured expert summary descriptions derived from expert consensus voting, rather than free-form narrative reasoning. The framework integrates signal-domain modeling of EEG with structured clinical language supervision. Each architectural component corresponds to a distinct analytical step commonly employed during EEG interpretation.

Figure~\ref{fig:framework_overview} presents a complete overview of the proposed pipeline. The methodological description follows the same left-to-right progression as the figure, beginning with EEG signal construction, followed by time--frequency transformation, dual-view representation learning, adaptive fusion, multimodal alignment, and joint optimization.

\subsection{Problem Formulation and Learning Objectives}

Let $\mathbf{X}_{\mathrm{raw}} \in \mathbb{R}^{C_{\mathrm{raw}} \times T_{\mathrm{raw}}}$ denote a multichannel EEG recording sampled at frequency $f_s$, where $C_{\mathrm{raw}}$ is the number of scalp electrodes and $T_{\mathrm{raw}}$ is the number of temporal samples. Each EEG segment corresponds to a clinically annotated event and is associated with a discrete label
\begin{equation}
y \in \mathcal{Y} = \{0,1,2,3,4,5\},
\label{eq:label_space}
\end{equation}
representing seizure, lateralized periodic discharges, generalized periodic discharges, lateralized rhythmic delta activity, generalized rhythmic delta activity, and other patterns.

Each EEG segment is also paired with a structured expert opinion description derived from expert consensus voting. The description is represented as a token sequence
\begin{equation}
\mathbf{t} = (w_1, w_2, \dots, w_L),
\label{eq:text_tokens}
\end{equation}
where $w_i$ denotes a token from a fixed vocabulary and $L$ is the maximum sequence length.

\begin{figure*}[t]
\centering
\includegraphics[width=\textwidth]{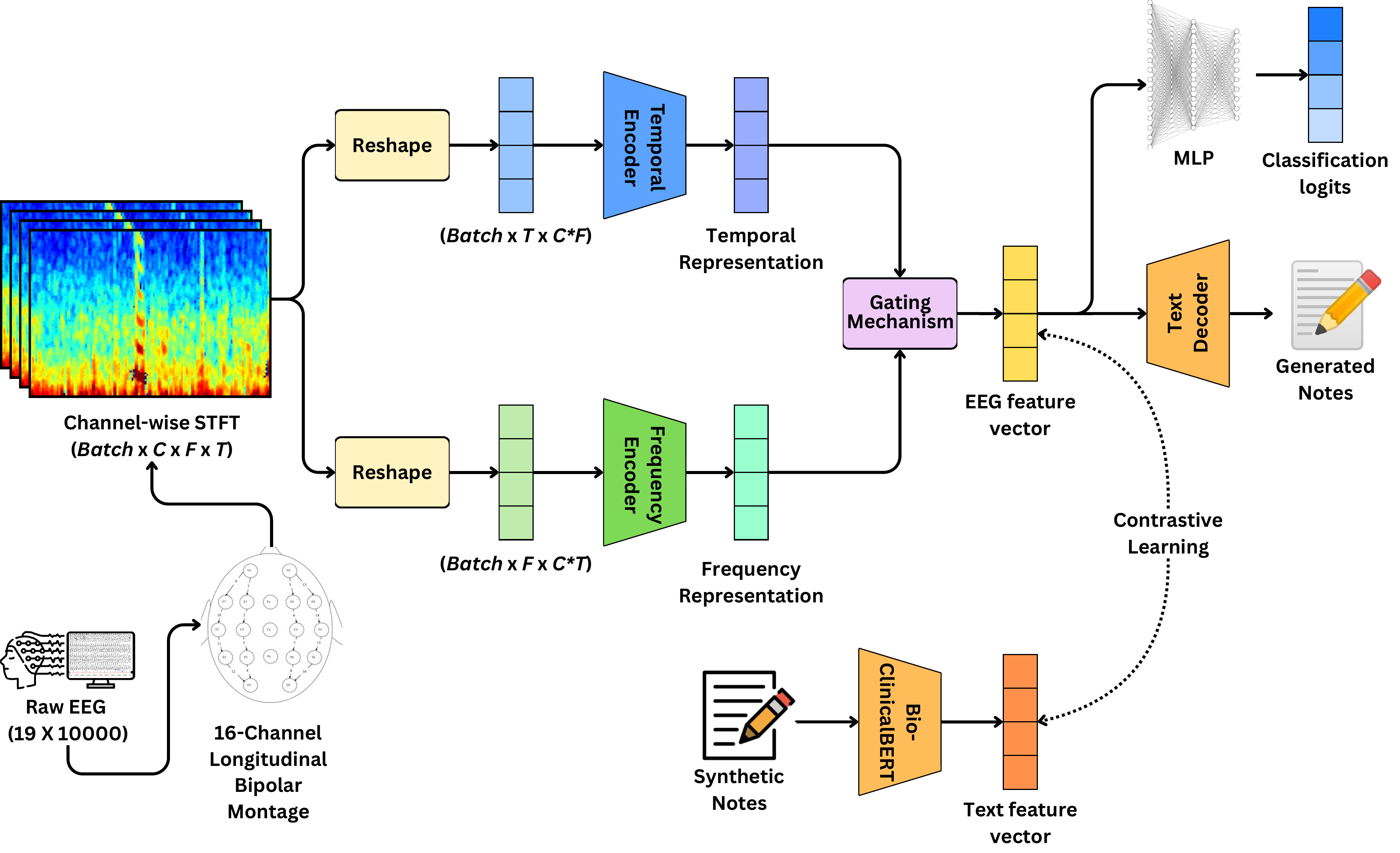}
\caption{Overview of the proposed multimodal EEG--text learning framework. Raw EEG signals are transformed into a longitudinal bipolar montage and segmented around clinically annotated offsets. Channel-wise short-time Fourier transform produces time--frequency representations, which are processed by two complementary transformer-based encoder branches emphasizing temporal-centric and frequency-centric dependencies. An adaptive gating mechanism fuses these representations into a unified EEG embedding. In parallel, structured clinical notes derived from expert consensus voting are encoded using a domain-specific language model. EEG and text embeddings are aligned through contrastive learning, while an auxiliary decoder reconstructs clinical text from EEG embeddings. The final EEG embedding is used for six-class harmful brain activity classification.}
\label{fig:framework_overview}
\end{figure*}

The learning objective jointly optimizes three related tasks: (i) classification of harmful brain activity from EEG, (ii) alignment between EEG representations and structured expert summaries, and (iii) reconstruction of expert summaries from EEG embeddings. These objectives reflect how EEG findings are categorized and communicated in clinical practice and motivate the multimodal structure shown in Fig.~\ref{fig:framework_overview}.

\subsection{EEG Input Construction and Time--Frequency Representation}

The first stage of the framework converts raw EEG recordings into a structured representation that preserves spatial relationships, temporal context, and spectral content. Errors or biases introduced at this stage propagate through downstream components, motivating explicit and physiologically grounded preprocessing.

\subsubsection{Longitudinal Bipolar Montage Construction}

Raw EEG channels are converted into a longitudinal bipolar montage (LBM), a standard representation used in clinical EEG review \cite{Tatum2014}. Bipolar derivations emphasize local voltage gradients and reduce sensitivity to reference-dependent artifacts.

Let $\mathbf{X}_{C_1}$ and $\mathbf{X}_{C_2}$ denote paired electrode signals forming longitudinal bipolar channel pairs. The bipolar EEG signal is computed as
\begin{equation}
\mathbf{X}_{\mathrm{bip}} = \mathbf{X}_{C_1} - \mathbf{X}_{C_2},
\label{eq:bipolar}
\end{equation}
yielding $\mathbf{X}_{\mathrm{bip}} \in \mathbb{R}^{C \times T}$. Each channel represents the potential difference between anatomically adjacent electrodes, improving spatial interpretability and robustness to global noise.

\begin{figure}[t]
\centering
\includegraphics[width=\columnwidth]{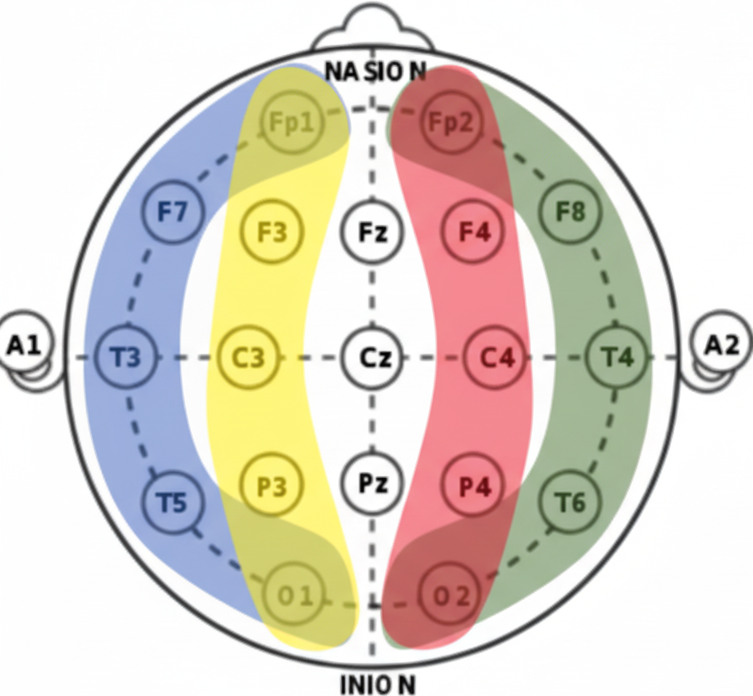}
\caption{Illustration of the longitudinal bipolar montage (LBM) used to derive EEG channels. Each bipolar channel represents the voltage difference between anatomically adjacent electrodes, emphasizing local spatial gradients and reducing reference-dependent artifacts.}
\label{fig:lbm_montage}
\end{figure}

\subsubsection{Window Extraction Around Labeled Offset}

Each EEG segment is extracted around a clinically annotated offset time $\tau$. Given a fixed window duration $T_w$ seconds, the extracted segment is defined as
\begin{equation}
\mathbf{X} = \mathbf{X}_{\mathrm{bip}}[:,\, t_0 : t_0 + T_w f_s],
\label{eq:window}
\end{equation}
where $t_0$ is chosen such that the window is centered on $\tau$. This ensures consistent temporal context across samples and aligns the input with expert annotation practice.

\subsubsection{Time--Frequency Transformation}

Many pathological EEG patterns are defined by rhythmicity, frequency content, and temporal regularity rather than absolute waveform morphology. To explicitly model these properties, a channel-wise short-time Fourier transform is applied.

For channel $c$, the STFT is defined as
\begin{equation}
\mathcal{S}_c(f,t) = \sum_{n=0}^{N-1} x_c[n]\, w[n-t]\, e^{-j2\pi fn},
\label{eq:stft}
\end{equation}
where $w(\cdot)$ is a Hann window. Retaining the magnitude yields a time--frequency tensor
\begin{equation}
\mathbf{Z} \in \mathbb{R}^{C \times F \times T_f}.
\label{eq:spectrogram}
\end{equation}

Time--frequency representations are standard in EEG analysis because periodic discharges and rhythmic delta activity are characterized by sustained spectral structure \cite{Oppenheim1999,Brunner2014}. The tensor $\mathbf{Z}$ serves as the shared input to the dual-view encoder.

\subsubsection{Spectrogram-Domain Data Augmentation}

To improve robustness, stochastic augmentations are applied to $\mathbf{Z}$ during training, including time masking, frequency masking, additive Gaussian noise, amplitude scaling, and temporal shifting. Augmentations are performed in the spectrogram domain to preserve physiologically meaningful structure while encouraging invariance to minor perturbations.

\subsection{Dual-View EEG Encoding}

The tensor $\mathbf{Z}$ contains intertwined spatial, temporal, and spectral dependencies. Modeling all dependencies jointly using a single encoder is computationally inefficient and dilutes inductive bias. The framework therefore decomposes $\mathbf{Z}$ into two complementary views prior to encoding.

\subsubsection{Temporal-Centric and Frequency-Centric Views}

A temporal-centric view is constructed as
\begin{equation}
\mathbf{Z}_{t} \in \mathbb{R}^{T_f \times (C F)},
\label{eq:temporal_view}
\end{equation}
emphasizing frequency-wise temporal evolution across channels.

A frequency-centric view is constructed as
\begin{equation}
\mathbf{Z}_{f} \in \mathbb{R}^{F \times (C T_f)},
\label{eq:frequency_view}
\end{equation}
emphasizing frequency-specific structure aggregated across channels and time.

This decomposition mirrors clinical EEG review, where temporal evolution and rhythmic structure are considered separately before integration.

\subsubsection{Linear Projection and Transformer Encoding}

Each view is projected into a latent space of dimension $d$:
\begin{equation}
\mathbf{H}_t = \mathbf{Z}_t \mathbf{W}_t, \quad
\mathbf{H}_f = \mathbf{Z}_f \mathbf{W}_f.
\label{eq:projection}
\end{equation}

Each projected sequence is processed by a stack of transformer encoder layers \cite{Vaswani2017}. Positional embeddings preserve ordering information, and self-attention enables long-range interactions:
\begin{equation}
\mathrm{Attn}(\mathbf{Q},\mathbf{K},\mathbf{V}) =
\mathrm{softmax}\left(\frac{\mathbf{Q}\mathbf{K}^\top}{\sqrt{d}}\right)\mathbf{V}.
\label{eq:attention}
\end{equation}

\begin{figure}[t]
\centering
\includegraphics[width=0.6\columnwidth]{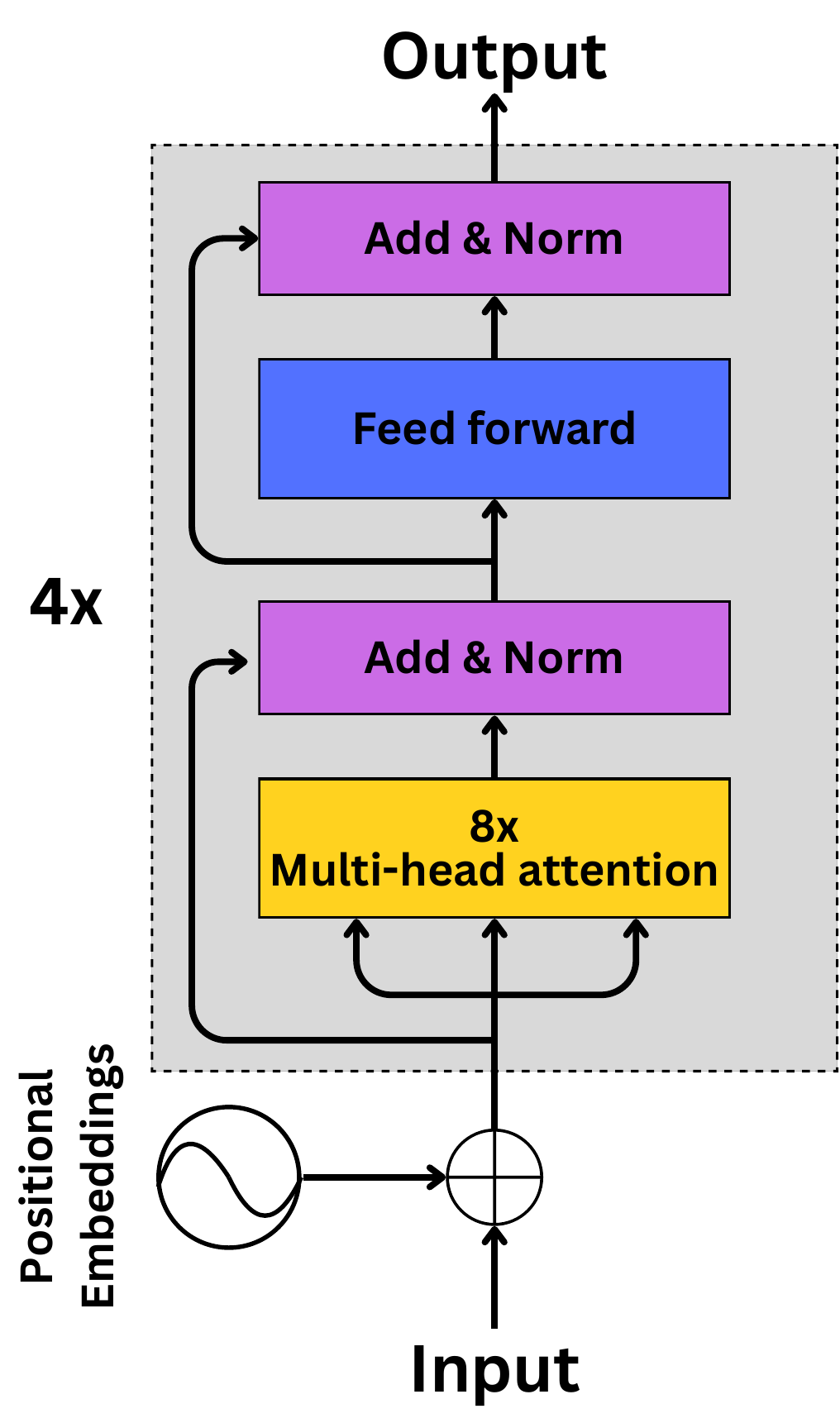}
\caption{Structure of the transformer encoder used in both temporal-centric and frequency-centric EEG branches. Each encoder consists of stacked multi-head self-attention layers with positional encoding, feed-forward networks, residual connections, and layer normalization.}
\label{fig:encoder_block}
\end{figure}

\subsubsection{Global Aggregation}

Token-wise outputs are aggregated using mean pooling:
\begin{equation}
\mathbf{h}_t = \frac{1}{C}\sum_{i=1}^{C} \mathbf{H}_t^{(i)}, \quad
\mathbf{h}_f = \frac{1}{F}\sum_{i=1}^{F} \mathbf{H}_f^{(i)}.
\label{eq:pooling}
\end{equation}

\subsection{Adaptive Gating and Feature Fusion}

The temporal and frequency embeddings are fused using a learnable, sample-specific gating mechanism. The concatenated embedding
\begin{equation}
\mathbf{h}_{\mathrm{cat}} = [\mathbf{h}_t ; \mathbf{h}_f]
\label{eq:concat}
\end{equation}
is mapped to fusion coefficients via
\begin{equation}
(\mu_t, \mu_f) = \mathrm{softmax}(\mathbf{W}_g \mathbf{h}_{\mathrm{cat}}).
\label{eq:gating}
\end{equation}

The final EEG embedding is
\begin{equation}
\mathbf{h}_{\mathrm{eeg}} = [\mu_t \mathbf{h}_t ; \mu_f \mathbf{h}_f].
\label{eq:eeg_embedding}
\end{equation}

The gating mechanism supports representational flexibility across heterogeneous EEG patterns and is not designed to guarantee monotonic improvements in classification accuracy.

\begin{figure*}[t]
\centering
\includegraphics[width=2\columnwidth]{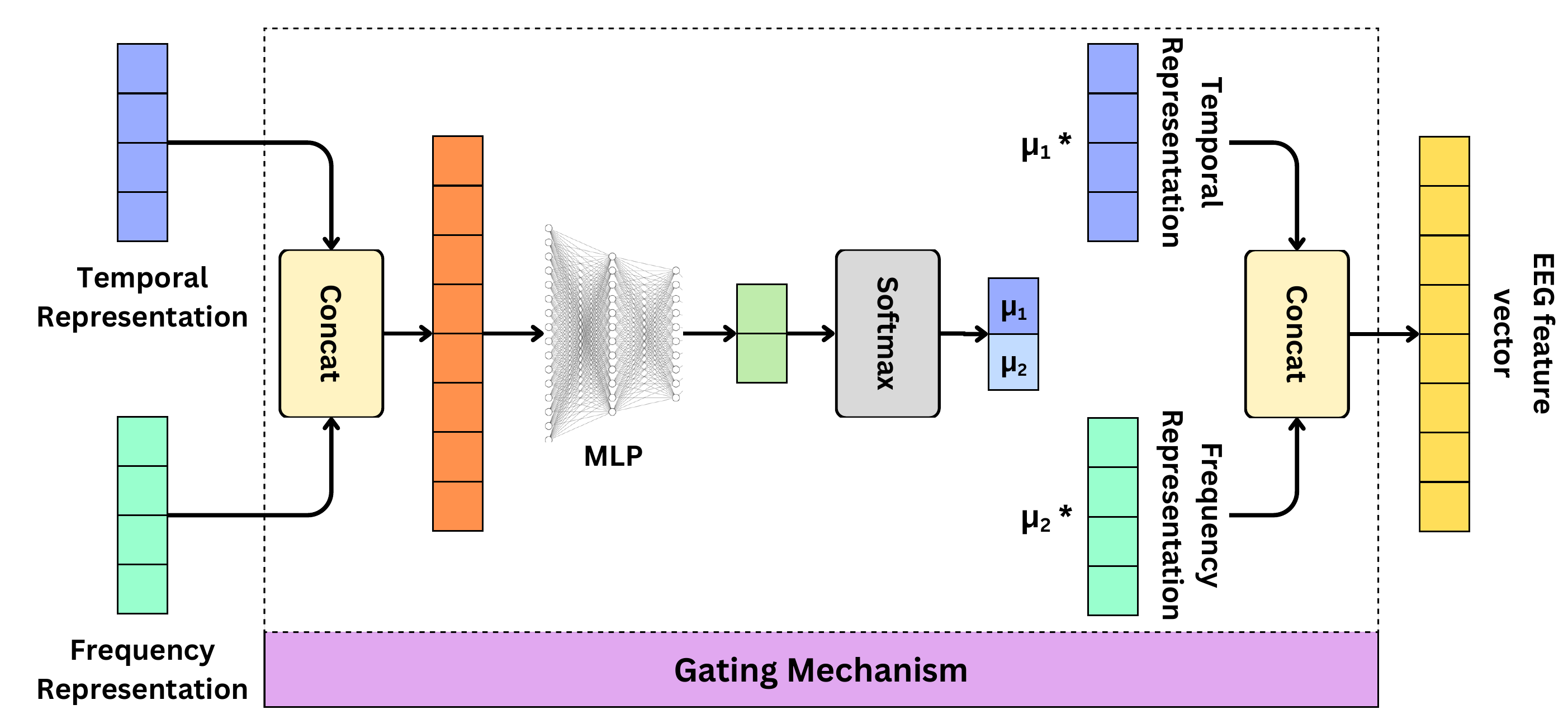}
\caption{Adaptive gating mechanism for fusing temporal-centric and frequency-centric EEG representations. Learnable coefficients determine the relative contribution of each branch to the final EEG embedding on a per-sample basis.}
\label{fig:gating_mechanism}
\end{figure*}

\subsection{Clinical Text Modeling and Multimodal Alignment}

Structured clinical descriptions are deterministically generated from expert vote distributions using standardized templates. These descriptions are encoded using Bio-ClinicalBERT, yielding
\begin{equation}
\mathbf{h}_{\mathrm{text}} = \mathbf{W}_p \,\mathrm{Bio\_ClinicalBERT}_{\mathrm{CLS}}(\mathbf{t}),
\label{eq:text_embedding}
\end{equation}
where only the final encoder layers are fine-tuned \cite{Alsentzer2019}.

The clinical descriptions are synthetic and follow a fixed format. Consequently, the text modality serves as a controlled semantic constraint on EEG representation learning rather than a source of linguistic variability. EEG and text embeddings are aligned using a symmetric contrastive loss \cite{Radford2021}.

\subsection{EEG-Conditioned Text Reconstruction}

An autoregressive transformer decoder reconstructs clinical text conditioned on $\mathbf{h}_{\mathrm{eeg}}$. The reconstruction loss is
\begin{equation}
\mathcal{L}_{\mathrm{rec}} = -\sum_{i=1}^{L} \log p(w_i \mid w_{<i}, \mathbf{h}_{\mathrm{eeg}}).
\label{eq:reconstruction}
\end{equation}

Text reconstruction is used as a diagnostic probe to assess whether EEG embeddings preserve clinically relevant descriptive information beyond class identity, and is not intended as a standalone report generation system.

\subsection{Classification and Joint Optimization}

EEG embeddings are classified using a multilayer perceptron:
\begin{equation}
\hat{y} = \mathrm{softmax}(\mathbf{W}_c \mathbf{h}_{\mathrm{eeg}}).
\label{eq:classification}
\end{equation}

The classification loss is
\begin{equation}
\mathcal{L}_{\mathrm{cls}} = -\log p(y \mid \mathbf{h}_{\mathrm{eeg}}).
\label{eq:cls_loss}
\end{equation}

All objectives are optimized jointly:
\begin{equation}
\mathcal{L} =
\lambda_1 \mathcal{L}_{\mathrm{cls}} +
\lambda_2 \mathcal{L}_{\mathrm{con}} +
\lambda_3 \mathcal{L}_{\mathrm{rec}},
\label{eq:total_loss}
\end{equation}
where $\lambda_i = \exp(\alpha_i)$ are learnable positive scalars \cite{Kendall2018}. The learnable weights balance gradient magnitudes across objectives and stabilize multi-task optimization.

\section{Experiments and Experimental Results}
\label{sec:experiments}

This section presents a quantitative evaluation of the proposed framework under strict subject-exclusive protocols. The analysis examines six-class classification accuracy, multimodal alignment, architectural sensitivity, parameter efficiency, and optimization behavior. Results are interpreted jointly across tables and figures to avoid isolated metric-driven conclusions. Because overall accuracy lies within a narrow operating range, effect sizes are reported using relative differences to highlight meaningful trends.

\subsection{Experimental Objectives}

The experimental evaluation addresses five objectives:

\begin{enumerate}
    \item Quantify six-class harmful brain activity classification performance.
    \item Evaluate EEG--text alignment using cross-modal retrieval metrics.
    \item Isolate the contribution of architectural components through ablation.
    \item Analyze performance versus model complexity trade-offs.
    \item Examine training stability and convergence behavior.
\end{enumerate}

\subsection{Dataset}
\label{subsec:dataset}

All experiments were performed using the \textit{HMS – Harmful Brain Activity Classification (HBAC)} dataset \cite{hms_kaggle}, developed jointly by Harvard Medical School, the Clinical Data Animation Center (CDAC), Persyst, and Jazz Pharmaceuticals. The dataset comprises 17{,}089 unique EEG recordings collected from approximately 1{,}950 patients under intensive and continuous neurophysiological monitoring. A total of 106{,}800 labelled 50-second EEG segments are provided at a uniform sampling frequency of 200\,Hz, each paired with a synchronized spectrogram covering a 10-minute temporal window centred on the labelled interval. The signals were recorded from 19 standard 10–20 system electrodes (Fp1, F3, C3, P3, F7, T3, T5, O1, Fz, Cz, Pz, Fp2, F4, C4, P4, F8, T4, T6, O2) and one electrocardiogram (EKG) reference channel, producing 20-channel bipolar recordings per sample. Expert epileptologists annotated the central 10 seconds of each 50-second window following the American Clinical Neurophysiology Society (ACNS) terminology, voting across six clinically relevant classes: Seizure (SZ), Lateralized Periodic Discharges (LPD), Generalized Periodic Discharges (GPD), Lateralized Rhythmic Delta Activity (LRDA), Generalized Rhythmic Delta Activity (GRDA), and Other. The majority vote determined the final consensus label (\texttt{expert\_consensus}), while individual vote counts for each category (\texttt{seizure\_vote}, \texttt{lpd\_vote}, \texttt{gpd\_vote}, \texttt{lrda\_vote}, \texttt{grda\_vote}, \texttt{other\_vote}) were retained to quantify inter-rater agreement. The final class distribution was moderately balanced, comprising 20{,}933 Seizure, 18{,}861 GRDA, 18{,}808 Other, 16{,}702 GPD, 16{,}640 LRDA, and 14{,}856 LPD samples. Each row in the metadata file \texttt{train.csv} specifies the EEG segment (\texttt{eeg\_id}, \texttt{eeg\_sub\_id}), its temporal offset within the full recording (\texttt{eeg\_label\_offset\_seconds}), and the corresponding patient identifier (\texttt{patient\_id}). Raw EEG signals are stored as individual \texttt{.parquet} files under \texttt{train\_eegs/}, with column names representing electrode channels and an additional EKG column.

\subsection{Experimental Setup}
\label{subsec:experimental_setup}

All experiments are conducted using a unified training and evaluation pipeline implemented in PyTorch. The configuration is designed to ensure architectural comparability across ablations, reproducibility of results, and stable optimization of high-capacity models.

\paragraph{Hardware Environment.}
Training and evaluation are performed on NVIDIA H100 GPUs. This accelerator supports transformer-based encoders operating on long EEG sequences and high-resolution time--frequency representations. All reported runtimes and convergence behavior correspond to this hardware configuration.

\paragraph{Software Stack.}
All models are implemented in PyTorch with CUDA-enabled backends. Automatic mixed-precision training is enabled to improve computational efficiency and reduce memory consumption while maintaining numerical stability \cite{Micikevicius2018}.

\paragraph{Training Protocol.}
All variants are trained for a fixed number of epochs. Early stopping is not applied. This design ensures that observed differences arise from architectural or objective-level changes rather than training duration. Optimization uses AdamW with decoupled weight decay \cite{Loshchilov2019}. Learning rate schedules and regularization parameters are identical across experiments unless explicitly modified by an ablation.

\paragraph{Batching and Data Loading.}
EEG segments are loaded using a fixed batch size across all runs. Data loading is parallelized to avoid input bottlenecks. Shuffling is applied at the subject level within the training split to preserve independence between training, validation, and test subjects.

\paragraph{Optimization Objectives.}
The full model optimizes a joint objective comprising classification, contrastive alignment, and text reconstruction losses. In ablation experiments, individual components are removed while all remaining objectives and weights are unchanged. This isolates the contribution of each objective.

\paragraph{Initialization and Reproducibility.}
All experiments use identical random seed configurations. Weight initialization schemes and optimizer hyperparameters are held constant. Evaluation uses the checkpoint obtained at the final training epoch, following established reproducibility guidelines \cite{Pineau2021}.

\paragraph{Evaluation Protocol.}
Performance is evaluated on a held-out test set. Test accuracy and retrieval metrics are computed using identical evaluation scripts across all variants. No test-time augmentation, ensembling, or calibration is applied.

\subsection{Summary of Quantitative Results}

Table~\ref{tab:ablation_results} summarizes results across all model variants. Metrics include test accuracy, cross-modal retrieval performance, trainable parameter count, validation loss, and training duration. This table serves as the numerical reference for subsequent analyses.

\begin{table*}[h]
\centering

\caption{Ablation study results under subject-exclusive evaluation. Metrics include test accuracy, EEG--text retrieval performance (Recall@K), number of trainable parameters, validation loss, and training epochs.}
\label{tab:ablation_results}

\begin{tabular}{lccccccc}
\toprule
\textbf{Model} & \textbf{Acc.} & \textbf{R@1} & \textbf{R@5} & \textbf{R@10} & \textbf{Params} & \textbf{Val. Loss} &  \\
\midrule
Full Model (Baseline)        & 0.9797 & 0.0694 & 0.2263 & 0.3390 & 34.38M & $2.3{\times}10^{-6}$\\
No Contrastive Loss          & 0.9784 & 0.0000 & 0.0018 & 0.0045 & 34.38M & $7.6{\times}10^{-7}$\\
No Text Reconstruction       & 0.9743 & 0.0654 & 0.2191 & 0.3210 & 16.34M & $2.2{\times}10^{-6}$ \\
EEG Only (No Text)           & 0.9766 & 0.0000 & 0.0000 & 0.0000 & 1.97M  & $6.2{\times}10^{-7}$ \\
Frozen Bio-ClinicalBERT      & 0.9793 & 0.0252 & 0.1073 & 0.1799 & 20.20M & $2.1{\times}10^{-6}$ \\
No Gating Mechanism          & 0.9811 & 0.0712 & 0.2259 & 0.3287 & 34.38M & $2.3{\times}10^{-6}$ \\
Reduced Transformer          & 0.9720 & 0.0667 & 0.2250 & 0.3291 & 33.85M & $2.2{\times}10^{-6}$  \\
No Data Augmentation         & 0.9725 & 0.0694 & 0.2209 & 0.3269 & 34.38M & $8.3{\times}10^{-6}$ \\
\bottomrule
\end{tabular}
\end{table*}

\subsection{Global View of Ablation Effects}

\begin{figure*}[h]
\centering
\includegraphics[width=\textwidth]{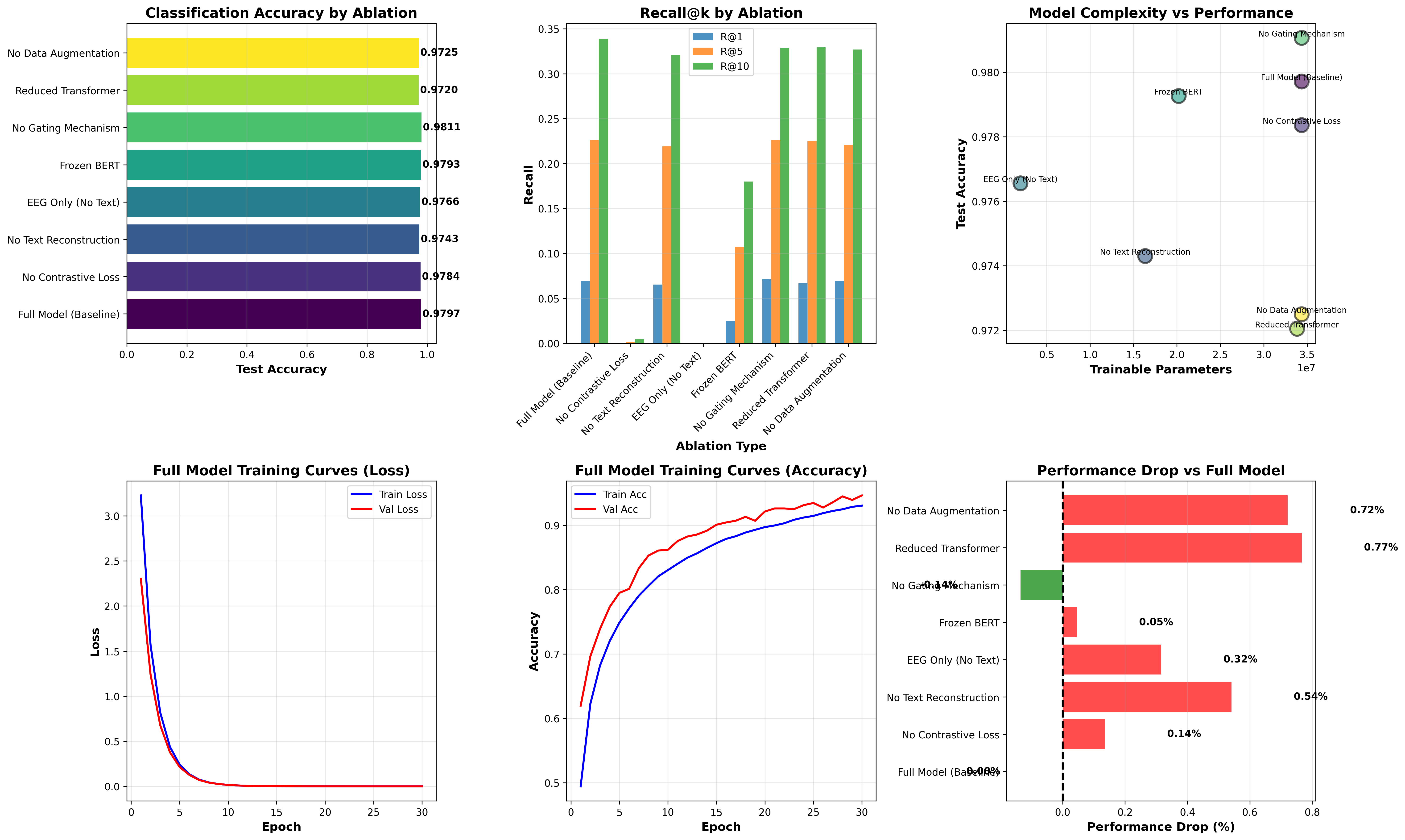}
\caption{Comprehensive ablation analysis showing (a) test accuracy, (b) EEG--text retrieval performance, (c) accuracy versus parameter count, (d) training and validation loss for the full model, (e) training and validation accuracy for the full model, and (f) relative performance change with respect to the baseline.}
\label{fig:comprehensive_ablation}
\end{figure*}

Figure~\ref{fig:comprehensive_ablation} provides a consolidated visualization of ablation outcomes. Each subfigure is analyzed in the following subsections.

\subsection{Classification Accuracy and Architectural Sensitivity}

As shown in Fig.~\ref{fig:comprehensive_ablation}(a), test accuracy ranges from 0.9720 to 0.9811 across variants, corresponding to a relative spread of approximately 0.93\%. The largest degradations are observed for the \emph{Reduced Transformer} and \emph{No Data Augmentation} variants, indicating that representational depth and exposure to diverse signal realizations dominate classification robustness.

Removing the gating mechanism yields a modest increase in accuracy. Given the narrow operating regime, this improvement corresponds to a small number of additional correct predictions and suggests that fixed fusion is sufficient to optimize decision boundaries for this dataset.

The EEG-only model achieves 0.9766 accuracy using only 1.97M parameters, corresponding to 99.27\% of baseline performance with approximately 5.7\% of the parameters.

\subsection{Cross-Modal Retrieval and Representation Structure}

Figure~\ref{fig:comprehensive_ablation}(b) reports EEG--text retrieval performance. The full model achieves Recall@10 of 0.3390. Removing the contrastive objective causes Recall@10 to collapse to 0.0045, a relative reduction of 98.7\%, despite minimal change in classification accuracy. This demonstrates a strong decoupling between discriminative accuracy and semantic alignment.

Removing text reconstruction yields a uniform reduction across Recall@K metrics, indicating that reconstruction refines alignment rather than defining it. Removing gating slightly reduces retrieval performance, revealing a trade-off between peak classification accuracy and semantic flexibility.

\subsection{Model Complexity and Efficiency Trade-offs}

Figure~\ref{fig:comprehensive_ablation}(c) illustrates diminishing returns from parameter scaling. Increasing parameters from 1.97M to 34.38M yields a relative accuracy gain of approximately 0.32\%.

The \emph{No Text Reconstruction} variant reduces parameters by over 50\% while preserving most classification and retrieval performance, indicating a favorable efficiency--performance balance. In contrast, reducing transformer depth degrades performance despite similar parameter counts, highlighting the importance of architectural structure over raw capacity.

\subsection{Training Dynamics and Optimization Stability}

Figures~\ref{fig:comprehensive_ablation}(d) and (e) show smooth and tightly coupled training and validation curves for the full model. Loss converges rapidly and validation accuracy tracks training accuracy closely, indicating stable optimization.

\begin{figure*}[h]
\centering
\includegraphics[width=\textwidth]{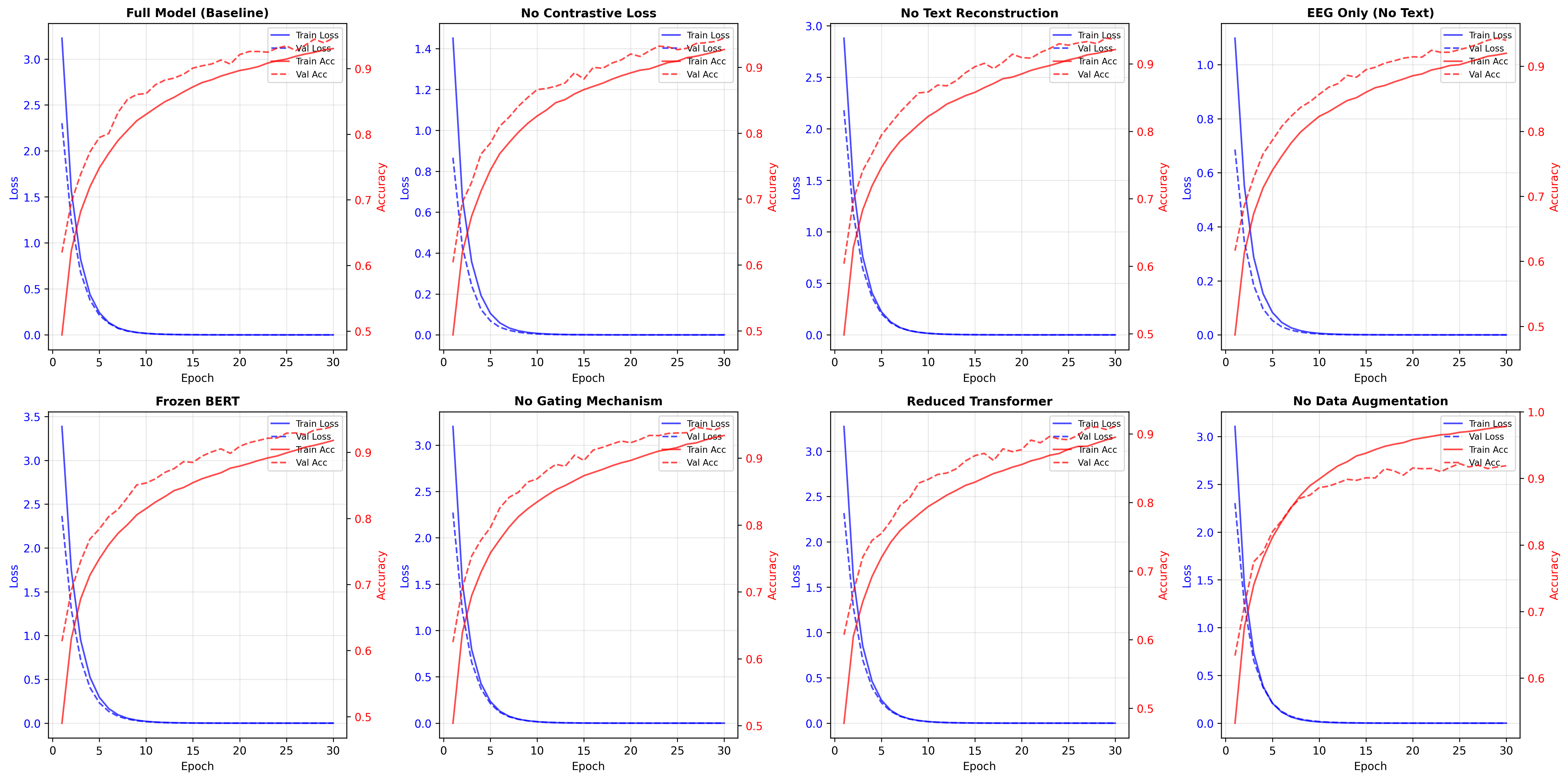}
\caption{Training and validation curves for all ablation variants. Models without multimodal supervision exhibit faster early convergence but increased validation variance.}
\label{fig:training_curves_all}
\end{figure*}

Figure~\ref{fig:training_curves_all} compares convergence across variants. Models without multimodal supervision converge faster initially but exhibit higher validation variance, indicating reduced optimization stability.

\subsection{Baseline Convergence Characteristics}

\begin{figure}[h]
\centering
\includegraphics[width=\columnwidth]{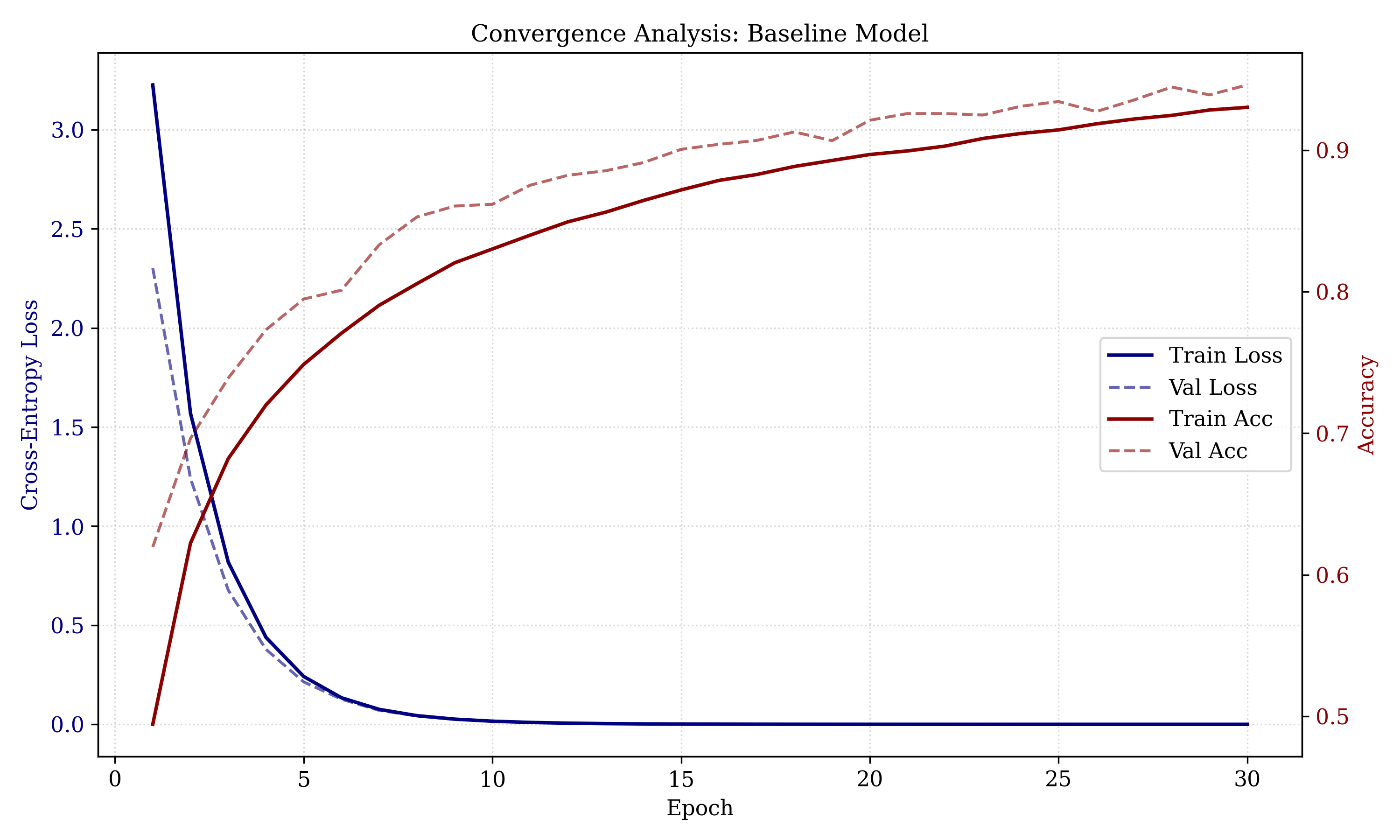}
\caption{Detailed convergence behavior of the full model, showing stable training and validation dynamics.}
\label{fig:baseline_convergence}
\end{figure}

Figure~\ref{fig:baseline_convergence} provides a detailed view of baseline convergence. Validation accuracy increases monotonically and reaches approximately 0.946 by epoch 30, while validation loss decreases to $2.3{\times}10^{-6}$, indicating well-conditioned optimization.

\subsection{Integrated Interpretation}

Across all analyses, classification accuracy, semantic alignment, and optimization stability are governed by partially independent factors. Encoder depth and augmentation dominate accuracy, contrastive learning governs semantic alignment, and auxiliary objectives stabilize optimization. Models with similar accuracy can differ substantially in representation quality, underscoring the need for multi-dimensional evaluation in clinically oriented EEG modeling.

\section{Discussion}
\label{sec:discussion}

This section interprets the experimental findings in the context of prior EEG modeling work and highlights their implications for representation learning in neurocritical care. The emphasis is on what the results reveal about architectural design, supervision strategies, and evaluation practices, rather than restating numerical outcomes or clinical motivation discussed earlier.

\paragraph{Beyond seizure-centric EEG modeling}

Most existing EEG deep learning studies are developed primarily for seizure detection, with non-seizure abnormalities either excluded or collapsed into coarse abnormal categories \cite{Schirrmeister2017,Lawhern2018,TjepkemaCloostermans2020,Mousavi2021TBME}. While this focus has enabled progress on seizure benchmarks, it limits insight into how models represent other clinically important EEG patterns that are prevalent in neurocritical care, including periodic discharges and rhythmic delta activity \cite{Foreman2012,Gaspard2014,Saab2020}.

By explicitly modeling six ACNS-defined harmful EEG patterns within a unified framework, this work enables direct comparison of how modern architectures encode distinct pathological dynamics. This broader scope responds to gaps identified in both clinical EEG literature and recent machine learning surveys, which note the limited coverage of non-seizure patterns and the lack of representation-level analysis in prior studies \cite{Hirsch2013,Roy2019,Shoeibi2021}.

\paragraph{Role of long-context modeling}

Transformer-based encoders exhibit improved robustness relative to reduced-context variants, consistent with prior observations that self-attention mechanisms better capture long-range temporal dependencies in EEG signals \cite{Song2022NeuroImage,Yuan2023TBME}. This behavior aligns with clinical understanding that rhythmic and periodic EEG patterns are defined by persistence and temporal regularity rather than isolated events \cite{Foreman2012,Struck2017}.

However, the results also show that architectural capacity alone does not induce clinically structured representations. Transformer encoders trained solely with discrete classification supervision achieve high accuracy but lack meaningful semantic organization. This finding echoes prior EEG transformer studies that report accuracy gains without examining representation structure, underscoring the need for supervision that reflects clinical semantics rather than label separability alone.

\paragraph{Impact of explicit EEG--text alignment}

Introducing an explicit EEG--text contrastive objective produces substantial changes in representation geometry and cross-modal retrieval performance, while having minimal impact on classification accuracy. Removing this objective causes near-complete collapse of alignment behavior despite preserved discriminative performance.

This decoupling mirrors evidence from vision--language and medical image--text models, where contrastive alignment is necessary to impose semantic structure on learned embeddings \cite{Radford2021,Jia2021,Irvin2019,Huang2023}. In the EEG domain, prior signal--language studies remain limited in scale and evaluation rigor \cite{Zhang2022MICCAI,Liu2023Sensors}. The present findings provide quantitative evidence that multimodal supervision alters EEG representations in ways that are not captured by accuracy-centric evaluation.

\paragraph{Text reconstruction as a representational constraint}

EEG-conditioned text reconstruction contributes modestly to classification performance but consistently improves representation compactness and optimization stability. This behavior indicates that reconstruction acts primarily as a constraint on representation completeness rather than as an independent task objective.

Similar patterns have been observed in multimodal medical AI, where generative objectives are used to probe whether latent representations preserve clinically relevant information beyond class identity \cite{Huang2023NatMI,Moor2023}. In this setting, successful reconstruction suggests that EEG embeddings retain sufficient descriptive information to support structured clinical summaries, complementing contrastive alignment without duplicating its effect.

\paragraph{Adaptive fusion and representational flexibility}

Adaptive fusion provides increased stability in alignment behavior and reduced sensitivity to architectural perturbations compared with static fusion, although peak classification accuracy differences are modest. This trade-off is consistent with prior work on adaptive and mixture-based architectures, which often prioritize robustness under heterogeneous inputs over marginal accuracy gains \cite{Shazeer2017,Zhou2021NeuroImage}.

Given that harmful EEG patterns emphasize different temporal and spectral characteristics, pattern-dependent fusion offers a principled mechanism to reflect this heterogeneity. In contrast, most existing multi-view EEG models rely on static fusion strategies \cite{Wang2022TBME,Ding2023TSception}, limiting their ability to adapt representation emphasis across pattern types.

\paragraph{Implications for EEG representation learning}

Taken together, the results indicate that classification accuracy, semantic alignment, and optimization stability are governed by partially independent factors. Encoder depth and data augmentation dominate discriminative performance, contrastive alignment governs semantic organization, and auxiliary objectives influence training dynamics. Models with similar accuracy can therefore differ substantially in representation quality and interpretability.

These findings support the use of multi-dimensional evaluation protocols for clinically oriented EEG modeling and caution against accuracy-only assessment. Explicitly aligning EEG representations with structured clinical descriptions provides a practical mechanism to probe and shape representation quality in ways that better reflect clinical interpretation.

\paragraph{Limitations and future directions}

Several limitations remain. All experiments are conducted on a single dataset with fixed acquisition characteristics, and cross-institutional generalization is not evaluated \cite{Roberts2021}. The use of structured, template-based text descriptions constrains linguistic variability and does not capture the full range of free-form EEG reporting. In addition, uncertainty calibration and probabilistic confidence modeling are not explicitly addressed.

Future work should examine cross-dataset transfer, uncertainty-aware objectives, and longitudinal EEG modeling. Recent advances in multimodal foundation models suggest potential pathways for scaling EEG--language alignment while maintaining clinical grounding \cite{Bommasani2021,Moor2023}.

\section{Conclusion}
\label{sec:conclusion}

This work examined how clinically harmful EEG activity can be modeled in a manner that reflects both discriminative performance and the descriptive structure used in neurocritical care practice. Rather than treating EEG interpretation as a purely signal-level classification problem, the proposed framework integrates signal-domain modeling with structured clinical language supervision to probe and constrain representation learning.

Empirical results demonstrate that high classification accuracy alone does not guarantee clinically meaningful representations. Architectures with nearly identical accuracy exhibit substantial differences in semantic alignment, optimization stability, and representation structure. Explicit EEG--text contrastive supervision plays a central role in shaping embedding geometry, while auxiliary reconstruction objectives and adaptive fusion mechanisms influence stability and representational flexibility without dominating accuracy outcomes.

By extending evaluation beyond accuracy-centric metrics, this study highlights the importance of representation-level analysis for clinically oriented EEG modeling. The findings suggest that multimodal supervision provides a practical mechanism to assess and guide EEG representations toward clinically interpretable structure, even when linguistic supervision is structured and constrained.

While the current evaluation is limited to a single dataset and template-based clinical descriptions, the framework establishes a foundation for studying EEG representation quality under supervision paradigms that better reflect how EEG findings are interpreted and communicated in practice. This work contributes to ongoing efforts to move EEG machine learning toward models that are not only accurate, but also aligned with clinical reasoning and reporting conventions.


\bibliographystyle{elsarticle-num} 
\bibliography{v2}

\end{document}